\documentclass[prl,aps,superscriptaddress,twocolumn,amssymb,showpacs]{revtex4-1}

\usepackage{color}
\usepackage{graphicx}
\usepackage{longtable}
\usepackage{epsfig}
\usepackage{dcolumn}
\usepackage{bm}
\usepackage{amssymb}

\begin{document}

\title{Entangled Spin-Orbital Phases in the d$^9$ Model}

\author{     Wojciech Brzezicki}
\affiliation{Marian Smoluchowski Institute of Physics, Jagellonian
             University, \\ Reymonta 4, PL-30059 Krak\'ow, Poland }

\author {    Andrzej M. Ole\'{s} }
\affiliation{Marian Smoluchowski Institute of Physics, Jagellonian
             University, \\ Reymonta 4, PL-30059 Krak\'ow, Poland }
\affiliation{Max-Planck-Institut f\"ur Festk\"orperforschung, Heisenbergstrasse 1, D-70569 Stuttgart, Germany }


\begin{abstract}
We investigate the phase diagrams of the spin-orbital $d^9$
Kugel-Khomskii model for a bilayer and a monolayer square lattice
using Bethe-Peierls-Weiss method.
For a bilayer we obtain valence bond phases with interlayer singlets,
with alternating planar singlets, and
two entangled spin-orbital (ESO) phases,
in addition to the antiferromagnetic and ferromagnetic order.
Possibility of such entangled phases in a monolayer is under
investigation at present.\\
{\it Published in: Acta Phys. Polon. A {\bf 121}, 1045 (2012).}
\end{abstract}


\maketitle

It has been shown that quantum fluctuations are enhanced near the
orbital degeneracy and could suppress long-range order in the
Kugel-Khomskii (KK) model \cite{Fei97}, called below the $d^9$
model. This model was introduced long ago for a perovskite
KCuF$_3$ \cite{Kug82}, a strongly correlated system with a single
hole within degenerate $e_g$ orbitals at each Cu$^{2+}$ ion. Kugel
and Khomskii showed that orbital order can be stabilized by a
purely electronic superexchange mechanism. This happens for
strongly frustrated orbital superexchange \cite{Ole10}, and
columnar Ising-type of order is obtained \cite{Cin10} in the
two-dimensional quantum compass model. This model exhibits
nontrivial symmetry properties which may be employed to perform
efficient calculations for square compass clusters \cite{Brz10}.

Orbital order occurs in a number of compounds with active orbital
degrees of freedom, where strong Coulomb interaction localizes
electrons (or holes) and gives rise to spin-orbital superexchange
\cite{Kha05}. When spin and orbital pseudospins couple to each
other, their order is usually complementary --- alternating
orbital (AO) order accompanies ferromagnetic (FM) spin order, and
ferro-orbital (FO) order coexists with antiferromagetic (AF) spin
order. However, the above Goodenough-Kanamori rules, see also
\cite{Ole10}, are not satisfied in cases when spin-orbital
entanglement (SOE) dominates \cite{Ole06}, as for instance in the
spin-orbital $d^1$ model on a triangular lattice \cite{Cha11}.

The spin-orbital superexchange KK model for Cu$^{2+}$ ($d^9$) ions
in KCuF$_3$ with $S=1/2$ spins and $e_g$ orbitals descibed by
$\tau=1/2$ pseudospin was derived from the degenerate Hubbard
Hamiltonian with hopping $t$, intraorbital Coulomb interaction $U$
and Hund's exchange $J_H$ \cite{Ole00}. It describes the
Heisenberg SU(2) spin interactions coupled to the orbital problem
by superexchange $J=4t^2/U$,
\begin{eqnarray} \label{hamik}
{\cal H}&=&-\frac{1}{2}J\!\!\sum_{\langle ij\rangle||\gamma}
\left\{\left(r_1\,\Pi^t_{\langle ij\rangle}+r_2\,\Pi^s_{\langle
ij\rangle}\right)
\left(\frac{1}{4}-\tau^{\gamma}_i\tau^{\gamma}_j\right)\right. \nonumber \\
&+& \left. r_3\,\Pi^s_{\langle
ij\rangle}\left(\frac{1}{2}-\tau^{\gamma}_i\right)
\left(\frac{1}{2}-\tau^{\gamma}_j\right)\right\}
-E_z\sum_{i}\tau_i^z\,.
\end{eqnarray}
where $\{r_1,r_2,r_3\}$ depend on $\eta\equiv J_H/U$ \cite{Ole00},
and $\gamma=a,b,c$ is the bond direction. In a bilayer two $ab$
planes are connected by interlayer bonds along the $c$ axis
\cite{Brz11} (a monolayer has only bonds within a single $ab$
plane). Here
\begin{equation}
\label{project} \Pi_{\langle ij\rangle}^{s}=\frac{1}{4}-{\bf
S}_i\cdot{\bf S}_{j}, \hskip .7cm \Pi_{\langle
ij\rangle}^{t}=\frac{3}{4}+{\bf S}_i\cdot{\bf S}_{j},
\end{equation}
are projection operators on a triplet (singlet) configuration on a
bond $\langle ij\rangle$, and $\tau^{\gamma}_i$ are the orbital
operators for bond direction $\gamma=a,b,c$. They are defined in
terms of Pauli matrices $\{\sigma^x_i,\sigma^z_i\}$ as follows:
\begin{eqnarray}
\tau^{a(b)}_i\equiv\frac{1}{4}\,(-\sigma^z_i\pm\sqrt{3}\sigma^x_i),
\hskip .5cm \tau^c_i=\frac{1}{2}\,\sigma^z_i.
\end{eqnarray}
Finally, $E_z$ is the crystal-field splitting which favors either
$x\equiv x^2-y^2$ (if $E_z>0$) or $z\equiv 3z^2-r^2$ (if $E_z<0$)
orbitals occupied by holes. Thus the model Eq. (1) depends on two
parameters: $E_z/J$ and $\eta$.

The spin-orbital model Eq. (\ref{hamik}) describes also CuO$_2$
planes in La$_2$CuO$_4$, where indeed $U\gg t$ and large
$E_z/J_H\simeq 0.27$ favors holes within $x$ orbitals
\cite{Ole00}. The superexchange between Cu$^{2+}$ ions $\sim
0.127$ eV reproduces there the experimental value. In this paper
we consider the model Eq. (\ref{hamik}) for K$_3$Cu$_2$F$_7$
bilayer compound where nearly degenerate $e_g$ orbitals are
expected. It has been shown that the magnetic state of
K$_3$Cu$_2$F$_7$ is described by interlayer valence bond (VB) phase
stabilized by FO$z$ order with $z$ orbitals occupied by holes
\cite{Man07}.

We show below that the bilayer spin-orbital $d^9$ model Eq.
(\ref{hamik}) describes a competition between different types of
spin-orbital order. Consider first $|E_z|\to\infty$, where
depending on the sign of the crystal field $E_z$ we get either
FO$z$ or FO$x$ configuration with $\langle\tau^c_i\rangle\equiv\pm
1/2$ and $\langle\tau^{a(b)}_i\rangle\equiv\mp 1/4$. After
inserting these values into Eq. (\ref{hamik}) one finds the
Heisenberg model describing either an AF bilayer ($E_z\to-\infty$)
or two independent AF planes ($E_z\to\infty$) as in La$_2$CuO$_4$.
In the limit of $\eta\to(1/3)^-$, the coefficient
$r_1=1/(1-3\eta)$ diverges and at large $\eta>0.26$ one finds
fully FM configuration with AO order.

The simplest approach is a single-site mean field (MF) approximation
applied to the model Eq. (\ref{hamik}). It excludes any spin
fluctuations so the spin projectors $\Pi^{t(s)}_{\langle ij\rangle}$
($\Pi^s_{\langle ij\rangle}$) can be replaced by their mean values,
where the dependence on the bond $\langle ij\rangle$ reduces to
direction $\gamma$ in phases with translationally invariant
magnetic order listed in Table~\ref{sord}: the $G$-AF phase, the
$C$-AF phase with AF planes and FM interplane bonds, the $A$-AF
phase with FM planes and AF interplane bonds and the FM phase.

\begin{figure}[t!]
    \includegraphics[width=8.5 cm]{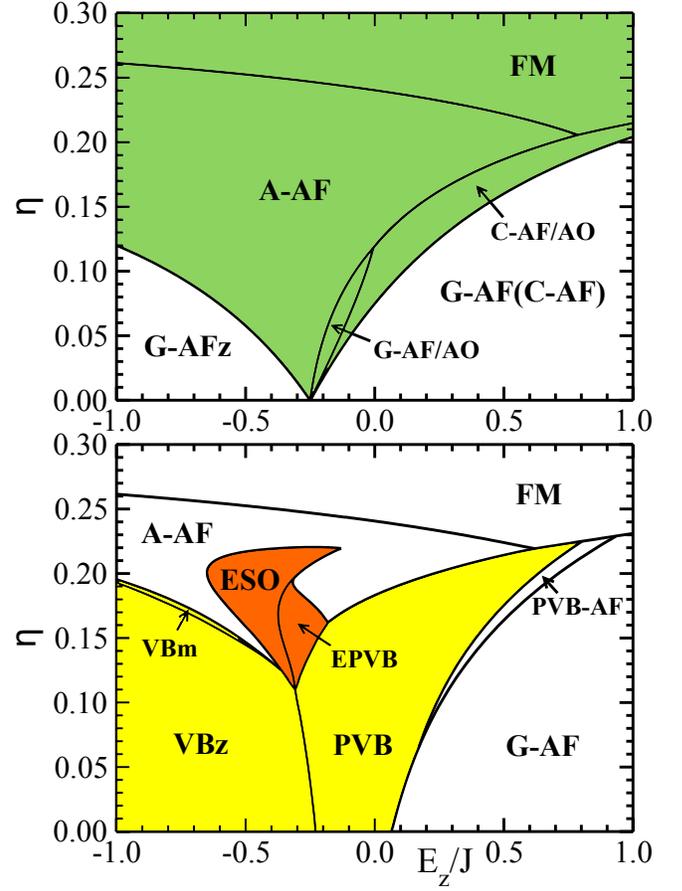}
\caption{Phase diagrams of the $d^9$ bilayer model (1) obtained
in: (a) a single-site MF, and (b) a cluster MF. In (a) shaded
(green) area indicates phases with nontrivial AO order while the
other ones have FO order.
In (b) light shaded (yellow) area marks singlet phases with spin
disorder and dark (orange) shading indicates phases with SO
entanglement. }
    \label{diags}
\end{figure}

In the orbital sector we apply then the MF decoupling for the
products $\{\tau_i^{\gamma}\tau_{i\pm\gamma}^{\gamma}\}$ along the
axis $\gamma$:
\begin{equation}
\tau_i^{\gamma}\tau_{i\pm\gamma}^{\gamma}
\simeq \langle \tau_i^{\gamma} \rangle \tau_{i\pm\gamma}^{\gamma}+
\tau_i^{\gamma}\langle \tau_{i\pm\gamma}^{\gamma}\rangle -
\langle \tau_i^{\gamma} \rangle\langle \tau_{i\pm\gamma}^{\gamma} \rangle.
\end{equation}
As order parameters we take $t^a\equiv\langle\tau_1^{a}\rangle$
and $t^c\equiv\langle\tau_1^{c}\rangle$ for a chosen site $i=1$
(which is sufficient in orbital sector as $t^b=-t^a-t^c$) and we
assume two orbital sublattices: each neighbor of the site $i$ is
rotated by $\pi/2$ in the $ab$ plane meaning that
$\langle\tau_{i+\gamma}^{a(b)}\rangle=t^{b(a)}$. The
self-consistency equations can be solved analytically (see Ref.
\cite{Brz11}) and the phase diagram of Fig. \ref{diags}(a) is
obtained by comparing the ground state energies for different
points in the $(E_z/J,\eta)$ plane. One finds two classes of
solutions: ($i$) uniform orbital configurations ($t^c=\pm 1/2$,
$t^{a(b)}=\mp 1/4$) for global FO order, and ($ii$) nontrivial AO
order with orbitals staggering from site to site in $ab$ planes.

\begin{table}[b!]
\caption{Mean values of triplet and singlet spin projection
operators (\ref{project}) for a bond $\langle ij\rangle$ in the
$ab$ plane and along the axis $c$ in magnetic phases with long
range order, see Fig. \ref{diags}.}
\begin{ruledtabular}
\begin{tabular}{cccccccc}
&   &   average  &  $G$-AF  &  $C$-AF  &  $A$-AF  &  FM  & \cr
\colrule & $ab$ plane & $\langle\Pi^t_{\langle ij\rangle}\rangle$
& 1/2 & 1/2 &  1  &  1  & \cr &
& $\langle\Pi^s_{\langle ij\rangle}\rangle$ & 1/2 & 1/2 &  0  &  0
& \cr
&  $c$ axis  & $\langle\Pi^t_{\langle ij\rangle}\rangle$ & 1/2 & 1
& 1/2 &  1  & \cr
&      & $\langle\Pi^s_{\langle ij\rangle}\rangle$ & 1/2 &  0  &
1/2 &  0  & \cr
\end{tabular}
\end{ruledtabular}
\label{sord}
\end{table}

For $\eta=0$ we have only two AF phases, see Fig. \ref{diags}(a):
$G$-AF$z$ for $E_z<-J/4$ and $G$-AF$x$ for $E_z>-J/4$, with
different FO orders involving $z$ or $x$ orbitals, respectively.
Because of the planar orbital configuration in the latter $G$-AF
phase one finds no interplane spin coupling and thus this phase is
degenerate with the $C$-AF one. For higher $\eta$ the number of
phases increases abruptly by three phases, all with AO
configurations: the $A$-AF, $G$-AF/AO and $C$-AF/AO phase.
Surprisingly, the AO version of the $G$-AF phase is connected
neither to FO$z$ nor to FO$x$ order in an antiferromagnet,
excluding the multicritical point at $(E_z/J,\eta)=(-0.25,0)$, and
disappears completely for $\eta\approx 0.118$. The $C$-AF/AO phase
stays on top of uniform $G$($C$)-AF phase, lifting their
degeneracy at relatively large $\eta$ and then gets replaced by
the FM phase which always coexists with AO order, so one can
conclude that the $G$-AF/$C$-AF degeneracy is most easily lifted
by turning on the orbital alternation. On the opposite side (for
$E_z<0$), the $G$-AF$z$ phase is completely surrounded by $A$-AF
phase with AO order. In the $A$-AF phase the AF correlations in
the $c$ direction survive despite the overall FM tendency when
$\eta$ grows. This follows from the orbitals' elongation in the
$c$ direction which stabilizes interplane singlets in a better
cluster MF approach, see below. Finally, the FM phase is favored
for any $E_z$ if only $\eta$ is sufficiently close to $1/3$, as
expected.

In a better cluster MF (or Bethe-Peierls-Weiss) approach,
introduced to capture the effects of quantum fluctuations, one
divides the bilayer square lattice into separate cubes containing
8 sites each and treats the bonds inside a cube exactly, and the
bonds connecting different cubes in MF. This approach has at least
three advantages over the single-site MF: ($i$) spins can
fluctuate, ($ii$) elementary cell can double, and ($iii$) we can have
independent spin-orbital order parameter. The MF leads in a cluster
to three order parameters: magnetic $\langle s\rangle\equiv S^z_1$,
orbital $t^{a(b)}$, and on-site SOE $r^{a(b)}\equiv\langle
S^z_1\tau^{a(b)}_1\rangle-st^{a(b)}$.

The self-consistency equations take rather complicated form (for
details see Ref. \cite{Brz11}) and can be solved only numerically
by time-consuming iterative Lanczos diagonalization of a cluster
combined with updating the MFs. In orbital sector apart from the
AO order described earlier, we consider configurations where
orbitals within a cluster break the symmetry between $a$ and $b$
directions but the neighboring clusters are rotated by $\pi/2$ in
the $ab$ plane, so globally the symmetry is preserved and the
elementary cell is doubled. In the spin sector we consider the
same configurations as in a single-site approach.

\begin{figure}[t!]
    \includegraphics[width=8.4 cm]{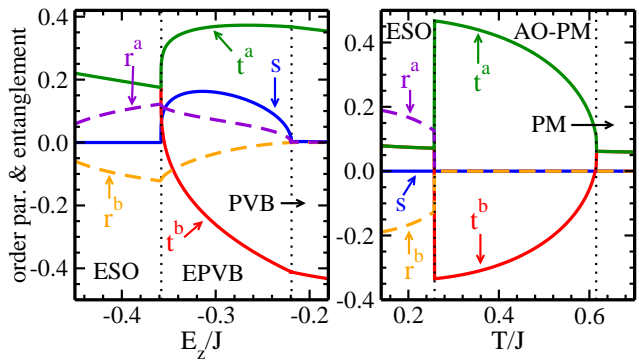}
    \caption{
Order parameters spin $s$, orbital $t^{a(b)}$ and on-site SOE
$r^{a(b)}$ for $\eta =0.15$ and for increasing: (a) $E_z$ in a
bilayer in the ESO, EPVB and PVB phase, for $T=0$; (b) $T$ in
a mono\-layer in the ESO, AO-PM and PM phase, for $E_z=-0.6J$.}
\label{enta}
\end{figure}

Including spin fluctuations in the cluster MF approach stabilizes
the $G$-AF phase with $x$ orbitals over the FM one ($E_z>0$) but
suppresses it when $z$ orbitals are filled by holes ($E_z<-J/4$),
and gives instead three singlet VB phases called:
PVB (plaquette VB), VB$z$ and VB$m$, see Fig. \ref{diags}(b).
VB$z$ phase replaces $G$-AF$z$ phase shown in Fig. \ref{diags}(a)
and involves interplane singlets accompanied by FO$z$
configuration. This phase was observed in K$_3$Cu$_2$F$_7$ by
Manaka {\it et al.\/} \cite{Man07} --- here we explain it for
realistic $\eta\simeq 0.14$. The VB$m$ phase is very similar to
VB$z$ but with slightly modified FO order by an AO component
increasing toward the $A$-AF phase. Transition from VB$m$ to VB$z$
is of the second order. In the PVB phase spin singlets are pointing
uniformly in $a$ or $b$ direction within the cluster and the
elementary cell is doubled.

A different class of phases involves SOE --- these are the ESO,
EPVB and PVB-AF phase. All of them exhibit SOE but only the ESO
and EPVB ones lie in the highly frustrated part of the phase
diagram and have large on-site entanglement $r^{a(b)}$, as shown
in Fig. \ref{enta}(a). The PVB-AF phase connects PVB and $G$-AF
phases by second order phase transitions and is characterized by
fast changes in orbital order and appearance of global
magnetization. The ESO phase has no magnetization and FO order is
here much weaker than in the VB$z$ phase. When $E_z$ grows, the
ESO phase does change continuously into the EPVB configuration,
being an entangled precursor of the PVB phase, with doubling of
the unit cell and finite AF order which vanishes smoothly
approaching the PVB phase. Additional calculations described in
\cite{Brz11} show that these entangled phases are absent if one
assumes that $\langle S^z_1\tau^{a(b)}_1\rangle$ factorizes, i.e.,
$r^{a(b)}=0$.

Using the same cluster MF approach as above one can easily study
the phase diagram of the KK model for a single layer at finite
temperature $T$. At T = 0 one finds
 the AF, FM, and PVB phases together with an
ESO phase between the AF and FM phases.
Turning on the thermal fluctuations we
have found that typically the orbital order is much
more robust than the magnetic one and the orbital
configuration compatible with lattice geometry
can greatly stabilize spin order. In Fig. \ref{enta}(b)
we present the thermal
evolution of the order parameters $\{s,t^{a(b)}\}$ and on-site SOE
parameter $r^{a(b)}$ in the ESO phase which melts and ends up as
an ordinary paramagnetic (PM) phase.
More details and the phase diagrams will be reported elsewhere.

Summarizing, we have shown that spin-orbital entanglement leads to
exotic types of order which are stabilized by quantum fluctuations
both in
bilayer and monolayer systems.. They emerge from highly
frustrated spin-orbital
superexchange and could be discovered only within a cluster mean
field approach.

\section*{Acknowledgments}

We acknowledge support by the Foundation for Polish Science (FNP)
and by the Polish National Science Center (NCN) under
Project No. N202~069639.

\end{document}